\documentclass[12pt]{article}
\usepackage{xcolor}
\usepackage{amssymb}
\usepackage{amsmath}
\usepackage{amsfonts}
\usepackage{graphicx}
\usepackage{mathrsfs}
\usepackage{comment}
\usepackage{cite}
\usepackage{authblk}
\usepackage{array}
\usepackage{subcaption}

\newcommand{\Slash}[1]{{\ooalign{\hfil#1\hfil\crcr\raise.167ex\hbox{/}}}}

\newcommand{\ket}[1]{ | {#1} \rangle }
\newcommand{\beq}{\begin{equation}}  \newcommand{\eeq}{\end{equation}}
\newcommand{\bef}{\begin{figure}}  \newcommand{\eef}{\end{figure}}
\newcommand{\bec}{\begin{center}}  \newcommand{\eec}{\end{center}}
\newcommand{\non}{\nonumber}  
\newcommand{\laq}[1]{\label{eq:#1}}  

\newcommand{\Eq}[1]{Eq.(\ref{eq:#1})}
\newcommand{\Eqs}[1]{Eqs.(\ref{eq:#1})}
\newcommand{\eq}[1]{(\ref{eq:#1})}
\newcommand{\Sec}[1]{Sec.\ref{chap:#1}}
\newcommand{\ab}[1]{\left|{#1}\right|}

\newcommand{\lac}[1]{\label{chap:#1}}

\def\({\left(}
\def\){\right)}

\def\diag{\mathop{\rm diag}\nolimits}

\def\O{\mathcal{O}}

\newcommand{\AND}{~{\rm and}~}
\newcommand{\EV}{ {\rm \, eV} }
\newcommand{\KEV}{ {\rm \, keV} }

\newcommand{\GEV}{ {\rm \, GeV} }

\def\a{\alpha}
\def\b{\beta}

\def\d{\delta}
\def\e{\epsilon}
\def\f{\phi}
\def\g{\gamma}
\def\h{\theta}
\def\k{\kappa}
\def\l{\lambda}
\def\m{\mu}
\def\n{\nu}

\def\s{\sigma}

\def\D{\Delta}

\def\tl{\tilde}
\def\*{\dagger}

\usepackage[normalem]{ulem}

\begin{document}
\begin{titlepage}
\begin{center}

\vspace{1.0cm}

{\Large\bf Undulators are ALP Factories
}
\vspace{1.0cm}

{\bf  Wen Yin$^{1,2}$ and Junya Yoshida$^{3}$}

\vspace{1.0cm}
{\em 
$^{1}${Department of Physics, Tokyo Metropolitan University, Tokyo 192-0397, Japan\\}
$^{2}${Department of Physics, Tohoku University, Sendai, 980-8578, Japan\\}
$^{3}${International Center for Synchrotron Radiation Innovation Smart, \\ Tohoku University, Sendai, 980-8578, Japan\\}
}

\vspace{1.0cm}
\abstract{ 
Axion-like particles (ALPs) are known to be produced through axion-photon conversion in the presence of a stationary external magnetic field. Devices such as undulators and wigglers, which are used widely for photon production, e.g., in synchrotron radiation facilities, inherently possess strong magnetic fields, making them potential sources for ALP production.
In this paper, we establish formalisms and formulas for studying ALP production in the ALP-photon-charged current system based on quantum field theory. We demonstrate that ALP production is inevitable in any undulator with the standard designs due to the electron Coulomb potential as well as a resonance effect depending on the ALP mass. In particular, ALPs are predominantly produced in a direction slightly misaligned with the photons' main direction. 
We propose placing detectors in the desired directions during operations of the originally planned experiments as an efficient approach to simultaneously probing ALPs.
The calculation methods and formulas developed in this study are applicable to ALP production from other environments and productions of other particles beyond the standard model relevant to synchrotron radiations. 
}

\end{center}
\end{titlepage}

\setcounter{footnote}{0}
\section{Introduction}

Axion-like particles (ALPs), pseudo-scalar bosons coupled to the photons, are leading candidates in physics beyond the standard model. 
Some basic properties of the ALPs, the mass, couplings, and how many kinds of ALPs exist, depend on the UV completions. 
For instance, many ALPs are predicted in string theory~\cite{Witten:1984dg, Svrcek:2006yi,Conlon:2006tq,Arvanitaki:2009fg,Acharya:2010zx, Marsh:2019bjr,Kitano:2021fdl,Kitano:2023mra}, a scenario referred to as axiverse. Also, simple quantum field theories (QFTs) contain ALPs~\cite{ DiLuzio:2017tjx, Lee:2018yak, Ardu:2020qmo,Yin:2020dfn}.

ALPs acquire masses through non-perturbative effects, and their mass range potentially extends from massless to extremely high values. In particular, ALPs can be much lighter than the electroweak scale, which is an interesting possibility. Some light ALPs that couple to gluons can solve the strong CP problem: the unnaturally small nucleon electric dipole moment in nature~\cite{Peccei:1977ur,Peccei:1977hh,Weinberg:1977ma,Wilczek:1977pj}. 
On the other hand, their small mass contributes to their longevity, making them viable candidates for dark matter in the Universe. The abundance of cold dark matter can be explained in a wide mass range depending on various simple mechanisms. 
The mechanisms connect the couplings, masses of the ALPs with the early Universe history, making the ALP dark matter probes of the early Universe.\footnote{For instance, an ALP can be produced through the misalignment mechanism~\cite{Preskill:1982cy,Abbott:1982af,Dine:1982ah}, which can probe inflation scales and cosmological histories by considering the inflationary Bunch-Davies distribution~\cite{Graham:2018jyp, Guth:2018hsa,  Ho:2019ayl, Matsui:2020wfx, Nakagawa:2022wwm, Murai:2023xjn}.
It can also be produced from cosmic string-wall networks if the Peccei-Quinn phase transition occurs~\cite{Klaer:2017ond, Gorghetto:2018myk, Kawasaki:2018bzv, Buschmann:2019icd, Buschmann:2021sdq,Gorghetto:2020qws,Saikawa:2024bta}.
It can be produced due to the inflaton decay with stimulated emission~\cite{Moroi:2020has,Choi:2023jxw} or through bubble/domain wall expansion \cite{Lee:2024oaz, Lee:2024xjb} (see also heavy dark matter production~\cite{Azatov:2021ifm,Baldes:2022oev, Azatov:2022tii,Azatov:2024crd,Ai:2024ikj}). It can even be produced thermally, similar to hot dark matter, due to stimulated emission~\cite{Yin:2023jjj}.
These scenarios provide connections between typical ALP dark matter masses and couplings with the cosmological history. 
}
Alternatively, ALPs can act as inflatons, where their small mass is related to the slow-roll conditions required for driving inflation~\cite{Daido:2017wwb,Daido:2017tbr,Takahashi:2019qmh,Takahashi:2023vhv,Narita:2023naj}. 
ALPs also affect the evolutions and lifetimes of astrophysical objects, and the couplings to standard model particles for light ALPs are constrained from them~\cite{Raffelt:1985nk,Raffelt:1987yu,Raffelt:1996wa, Ayala:2014pea, Straniero:2015nvc, Giannotti:2015kwo, Carenza:2020zil}. Additionally, ALP helioscopes provide important constraints on ALPs produced by the Sun~\cite{Anastassopoulos:2017ftl} (see also future prospects by IAXO~\cite{Irastorza:2011gs, Armengaud:2014gea, Armengaud:2019uso, Abeln:2020ywv}).
However, some light ALPs can only be probed on Earth due to matter effects~\cite{Masso:2005ym,Jaeckel:2006xm,Brax:2007ak}. In addition, ALP production in ground-based experiments is not affected by astrophysical systematic uncertainties.

 A notable feature of ALPs is their ability to undergo axion-photon conversion. In the presence of a stationary external magnetic field within a finite volume, breaking Lorentz invariance,  massless photons can convert into massive ALPs and vice versa. 
This phenomenon has led to the proposal of experimental setups involving two magnetic field regions separated by a photon shield positioned in front of a photon beam source for detecting ALPs\cite{Sikivie:1983ip,Anselm:1985obz,VanBibber:1987rq}. The setup is referred to as light-shining-through-a-wall (LSW) experiment~\cite{Abel:2006qt,Arias:2010bh,Redondo:2010dp,Bahre:2013ywa,Ortiz:2020tgs,Seong:2023ran,Hoof:2024gfk}. Already various experimental results are reported \cite{Ehret:2010mh,Betz:2013dza,DellaValle:2015xxa,Inada:2016jzh,OSQAR:2015qdv, Halliday:2024lca} (see also \cite{SAPPHIRES:2022bqg}). Interestingly, LSW experiments can probe all ALPs with masses lighter than a specific threshold determined by the experimental configuration, provided the ALP-photon coupling is sufficiently strong and/or the number of ALPs is large enough.

One type of photon source is known as an undulator or wiggler. 
Undulators are sophisticated devices primarily used in synchrotron radiation facilities and are essential for a wide range of applications in science and technology, including materials science, molecular biology, the operation of free-electron lasers, and colliders, to produce highly directional and brilliant synchrotron radiation. 
They generate, or aim to generate, high-brightness beams of light ranging from infrared to hard X-rays, extending up to the $\gamma$-ray energy range.
The operation of undulators involves passing a relativistic electron beam through a spatially periodic magnetic structure, causing the electrons to undergo sinusoidal motion and emit radiation at specific wavelengths (See Fig.\ref{fig:concept}). 

Inside the undulator, both photons and a strong magnetic field are present. This raises the question of whether the undulator itself could serve as an ALP factory. 
This question is particularly relevant to the so-called third-generation and later synchrotron light sources equipped with multiple undulators worldwide.
If undulators themselves produce ALPs, this would imply that more than half of the LSW experimental setup—specifically, the installation of experimental devices that produce ALPs and the wall that shields the photons—is already in place.
However, estimating the ALP production by the undulator is not an easy task because the electric and magnetic fields inside the undulator depend non-trivially on time and space due to electron propagation.

In this paper, we carefully examine whether undulators can independently produce ALPs, using first-principles calculations in a system involving electromagnetic field, electric current, and an ALP, both analytically and numerically. 
For light ALPs, a dominant source is the Coulomb potential of the electron in conjunction with the external magnetic field, which does not contribute to the original synchrotron radiation. When the ALP has the preferred mass determined by the design of the undulator, resonant production can occur. 
Various results are consistent with a numerical estimation by using background electromagnetic fields and symmetry argument.
This indicates that any undulator is an ALP factory, continuously producing ALPs as long as the facility is operational. 
Placing a proper detector at the desired position can search for ALPs, giving the potential limits even better than the existing LSW limit in the resonant case thanks to the long `experimental time' equal to the facility running time. Our calculation methods can be generalized to other cases. For example, we can easily check that the ALP production from electrons in a circular collider is suppressed, and in some cases, similar analysis works in producing ALPs from astrophysical objects.  

The phenomena we discuss here differs from the `axion bremsstrahlung' discussed in the literature, such as in \cite{Tsai:1986tx}, where the axion production is via the axion-electron coupling, and it is different mechansim from ours. In our work, ALPs are produced solely due to axion-photon coupling, with no axion-electron coupling introduced. 

In \Sec{1}, we review synchrotron radiation in an undulator using standard methods and perturbation theory in QFT. 
In \Sec{ALP}, 
we examine ALP production from undulators and highlight the significance of the Coulomb potential.
The last section is devoted to the conclusions and discussions. 
The consistency of the main results with  the symmetry arguments and the numerical estimations by using background electromagnetic fields 
can be found in  Appendices.~\ref{chap:symmetry} and \ref{chap:nume}, respectively.

Throughout the paper we use the convention of the metric $\eta_{\m\n}\equiv\diag(1,-1,-1,-1)$, with $\m,\n=0,x,y,z$. A four-vector is denoted in the form $V^\m\equiv (V^0, \vec{V}) =(V^0, V^x,V^y,V^z).$
\section{Photons from Undulators}
\lac{1}
In this section, we review the usual synchrotron radiation from a single electron passing through an undulator.
The motion of the electron is discussed in \Sec{motion}. 
We employ two methods to estimate the photon flux from the undulator, both of which will agree with each other. 
The first method, which is the conventional one in the synchrotron radiation study, involves using the solution of Maxwell's equations to demonstrate synchrotron radiation resulting from the electron's acceleration (\Sec{Maxwell}).
The second method employs a field-theoretic approach, which will be used in the ALP production in \Sec{ALP} (\Sec{QFT}).

\subsection{Electron Motion in Undulators}
\lac{motion}

In an undulator, a moving electron is accelerated by a stationary but spatially oscillating magnetic field.
For simplicity and clarity, we consider an electron moving primarily in the $z$-direction, with the magnetic field causing undulations in the $x$- and $y$-direction. 
The position of the electron at time $t_e$ is denoted by $\vec{r}_e(t_e).$ The magnetic field is assumed to have components that only depend on $z$:
\beq
\vec{B}_{\rm ext}(z)= \(B_{{\rm ext}}^x(z),B_{{\rm ext}}^y(z),0\).
\eeq
This magnetic field strength is assumed to be non-vanishing in the regime $z=[0,L].$ 
We consider the Lorentz factor of the electron, $\g$, to be much greater than unity $\g\gg1$ and use $1/\g$ as an expansion parameter.
At $t_e=0$,  the $z$-component of the electron's position $r_{e}^z=0$.

The equation of motion for an electron in the magnetic field is given by
\beq 
\frac{d}{dt_e} \(\gamma m_e \vec\beta(t_e)\) = -e \vec{\beta}(t_e)\times \vec{B}_{\rm ext}(\vec{r}_e(t_e)).
\eeq 
Here, $e$ is the photon coupling, and $m_e$  is the electron mass. 
We neglect the friction due to the photon radiation because the Lorentz factor is large, and $L$ is not so long. 
Since a stationary magnetic field does not work to increase the kinetic energy of the electron, the Lorentz factor $\gamma$ remains constant.
Therefore,
$
\dot{\vec{\beta}}(t_e)=  -\frac{e}{\gamma m_e} \vec{\beta}(t_e)\times \vec{B}_{\rm ext}(\vec{r}_e(t_e)).
$

Giving that the $z$-component of $B_{\rm ext}$ is zero, we obtain 
\beq
\dot{\beta}^x =\frac{e}{\g m_e} \beta^z B_{{\rm ext}}^y, \dot{\beta}^y= -\frac{e}{\g m_e}\beta^z B_{{\rm ext}}^x
\eeq
Throughout this paper, we neglect the time variance of $\beta^z$.\footnote{Taking into account the time variance gives a higher order contribution in the $K$ expansion than those we consider in this paper.} 
 We use the standard form for the magnetic field of an undulator:
\beq
\vec{B}_{\rm ext}=B_0\{ \cos[k \beta^z t_e], \k \sin[k \beta^z t_e+\f] , 0\}.
\eeq
Here, $k$ denotes the wave number of the undulator, a characteristic property of the undulator. 
Without loss of generality, we assume that the parameters $\k \AND \f$, satisfy $0\leq \k\leq 1$ and $0\leq \f<2\pi$, which is set by the design of the undulator. 
For instance, a helical undulator has $\k=1,\f=0$, while a linear undulator has $\k=0$.

The solution to the equation of motion is straightforward 
\beq
\beta^x=-\frac{e B_0 \k }{\g k m_e} \cos (k \beta^z t_e+\f), ~~~~ \beta^y = -\frac{e B_0}{\g k m_e} \sin (k\beta^z t_e),
\eeq
by omitting constant terms. 
We obtain
\begin{align}
\vec{\beta}&\simeq \(-\frac{K\k}{\gamma} \cos[k \beta^z t_e+\f],- \frac{K}{\gamma} \sin[k\beta^z t_e], \beta^z\), \non \\
\AND \laq{sol}
\vec{r}_e&\simeq \(-\frac{K\k}{\gamma  k \beta^z} \sin[k\beta^z t_e+\f],\frac{K }{\gamma  k  \beta^z}\cos [k \beta^z t_e],t_e \beta^z\)
\end{align}
with 
\beq
\beta^z\simeq \sqrt{1-\gamma^{-2}\(1+\frac{1}{2}(1+\k^2)K^2\)},
\eeq
 where we approximate the perpendicular velocity by the time average.
 Here, we defined the $K$-parameter,
\beq
\laq{Kpara}
K\equiv \frac{B_0e}{k m_e}.
\eeq
 If $\k=1,\f=0$, i.e., a helical undulator, 
the previous forms \eq{sol} are exact.

     \begin{figure}[t!]
    \begin{center}  
    \includegraphics[width=165mm]{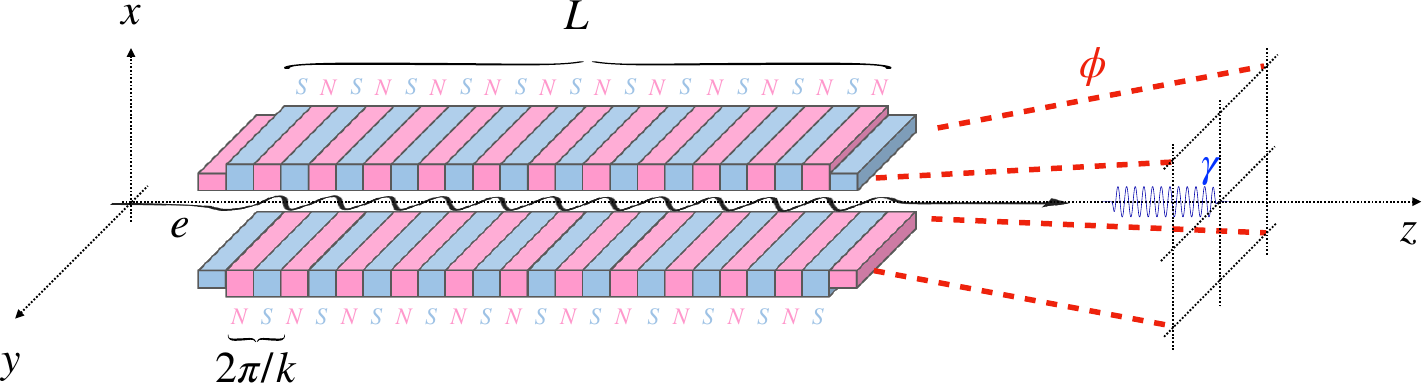}
    \end{center}
    \caption{ A conceptual design of a linear undulator, corresponding to $\k=0$.
    The energetic electron injected in $z$-direction undulates (black solid line) due to the magnetic field by the magnets placed along the $y$, $z$ plane (light blue and red blocks denote the south pole and north pole, respectively). The correspondences of the parameters $L$ and $k$  are also indicated.
    Due to the electron undulation, photons are preferably radiated along the $z$-axis  (blue wavy line denotes the magnetic field component of the photons). 
     In this paper, we show that ALPs are also produced by any undulator, with the preferred direction depending on the design and the mass of the ALP. However, the $z$-axis is always disfavored (red dashed lines denote the favored direction for the resonant ALP production for the linear undulator, see \Eq{angdep2} and \Sec{reso}). 
     }
    \label{fig:concept}
    \end{figure}

For a typical undulator, \beq \laq{smallK} K\lesssim 1
\eeq
which is the focus of this paper. 
In this case, the resulting photon wave is primarily aligned along the $z$-axis, closely following the shape of the electron trajectory with the Doppler effect. The conceptual design of a linear undulator with $\k=0$ is shown in Fig.~\ref{fig:concept}.
For illustrative purposes, we shifted the origin of the $z$-axis in the figure from the convention for our estimation.

When $K\gg 1$, which is called Wiggler, many approximations in this paper are not valid.
A Wiggler is known to produce photons with continuous spectrum, in contrast to the undulator, which produces photons with spectra in narrow lines.

\subsection{Synchrotron Radiation from Maxwell's Equations}
\lac{Maxwell}
One method to estimate the photon flux from the undulator is by solving Maxwell's equations, given the motion of the electron.
Our starting point is a general equation for the radiated electric field at the position $\vec{x}=\vec{R}+\vec{r}_{e}(t_e)$ due to a previously accelerating electron located at $\vec{r}_e(t_e)$ which can be straightforwardly derived using the delayed potential of the gauge field~\cite{Jackson:1998nia}: 
\beq\laq{ele}
-\vec{E}=\frac{e(1-\b^2)}{4\pi R^2 b^3}\(\frac{\vec{R}}{R}-\vec \beta\)+\frac{e}{4\pi R b^3}\frac{\vec{R}}{R}\times \left\{\(\frac{\vec{R}}{R}-\vec{\beta} \)\times \dot{\vec{\beta}}\right\},
\eeq
where 
\beq 
b\equiv 1-\frac{\vec{R}}{R}\cdot\vec{\beta}=1-\beta \cos{\theta},\AND t_e \equiv t-R \laq{terela}
\eeq
Here, $\theta$ is the angle between $\vec{R}$ and $\vec\beta(t_e)$, which is also the angle from the $z$-axis when $K/\g\ll1$.
 We defined $X\equiv |\vec{X}|\AND \dot X\equiv \frac{d}{d t_e} X $ here and in the following.  
The minus sign on the left-hand side indicates that the electron carries a charge of $-1$. 
The first term of \Eq{ele} represents the Coulomb potential, and the second term corresponds to photon radiation.
Since the second term is proportional to $\dot{\vec\beta} \propto K$, this radiation term vanishes as $K\to 0$. 
In contrast, the Coulomb potential term does not vanish in this limit. 
Indeed, from the delayed potential for the gauge field, one can verify that the magnetic field is given by $\vec{R}/R \times $ the second term, which serves as the solution to the electromagnetic wave.
Using the second term of \Eq{ele} and the magnetic field, we can estimate the Poynting vector, which is given by
\beq
\vec{S}= e^2 \frac{\vec{a}^2}{16\pi^2 R^2} \frac{b^2-\g^{-2}\cos^2\theta'}{b^6} \frac{\vec{R}}{R}.
\eeq
Here, $\vec{a}$ represents the acceleration of the electron, defined as $\vec{a}\equiv \frac{d}{dt_R}\vec{\beta}$, and $\theta'$ is the angle between $\vec{a}$ and $\vec R$.

By noting that $\vec{S}$ denotes the energy flux per unit time at a fixed position in terms of $t$, the energy radiated from electron per unit time in terms of $t_e$ is given by \beq
\frac{d P}{d \Omega} = R^2\vec{S} \cdot \frac{\vec{R}}{R} \frac{dt}{dt_R}  = \frac{e^2 \vec a^2}{16\pi^2 } \frac{1}{\(1-\beta \cos \theta \)^3} \(1- \frac{\g^{-2}\cos^2\theta'}{(1-\beta \cos\theta)^2}\).
\eeq
Substituting the solution of our equation of motion~\eq{sol}, we can calculate the power for the photon radiation. 
For instance with $\k=1, \phi=0$, the vector $\vec{a}$ rotates in the $x$- and $y$-direction. 
By noting that the time average of $\cos^2\theta'$ is $\frac{1}{2}\sin^2 \theta$, 
we obtain 
\beq
\laq{photonflux}
\frac{d P}{d \Omega}= \frac{e^2 K^2 k^2 \beta^2 }{16\pi^2 \g^2}\frac{1}{\(1-\beta \cos\theta\)^3}\(1-\frac{\g^{-2}\sin^2\theta}{2\(1-\beta \cos\theta\)^2}\)
\eeq
When $|\theta|\ll 1/\g$, i.e., when the photon energy is emitted along the $z$-direction, we get
\beq
 \frac{d P}{d \Omega}\simeq \frac{e^2 K^2 k^2 \beta^2 }{16\pi^2 \g^2} \frac{\g^6}{8(1+K^2)^3}.
\eeq
This scales with $\g^4$, representing the dominant contribution to the radiation for $K\lesssim 1$.

\subsection{Synchrotron Radiation for Particle Theorists}
\lac{QFT}

     \begin{figure}[t!]
    \begin{center}  
    \includegraphics[width=40mm]{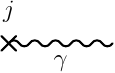}
    \hspace{30mm}
        \includegraphics[width=40mm]{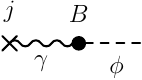}
    \end{center}
    \caption{ Feynman diagrams of QFT calculations for synchrotron radiation (left panel) and ALP production (right panel, see \Sec{ALP}).}
    \label{fig:Feynman}
    \end{figure}

For convenience in later discussions, let us provide a field-theoretical estimation of the radiation probability.
Consider the Lagrangian
\beq
\laq{Lag}
{\cal L}=-\frac{1}{4}F^{\m\n}F_{\m\n}-A^\m j_\m ,~~~ j_\m= -e v_\m(t) \delta^3(\vec x-\vec{r}_e[t]).
\eeq
with $v^\m=(1,\vec\b)$ being the velocity vector of electron extended to 4D, and $F_{\m\n}$  is the photon field strength tensor.
To study the photon production, we use perturbation theory to estimate the Feynman diagram shown in the left panel of Fig. \ref{fig:Feynman}. 
The amplitude is given by 
\beq
\laq{amp}
\langle{\g,\e, k_\g|0_j\rangle}=-\int d^4x  \frac{i \e^\m}{\sqrt{2w_{\g}}}j_\m e^{i (w_\g t-\vec{k}_\g \vec{x})  }
= -\frac{i}{\sqrt{2 w_{\g}}}\e^\m\tl j_\m(k_\g)
\eeq
Here, $\e$ is the polarization vector of the photon; $\vec{k}_\g$ denotes the photon momentum, and $w_\g$ denotes the energy with
$w_\g=k_\g$;
$0_X$ denotes the vacuum state with the presence of the background $X$, and it will take $X=j,E, {\rm and/or}~ B$, which is the current, electric field, and/or magnetic field in this paper; 
we have defined
\beq
\laq{tljm}
\tl j_\m (q)\equiv - e \int_{-\infty}^{\infty} dt      v_\m e^{i( q^0 t-\vec q  \cdot \vec r_{e}[t])}.
\eeq
By neglecting $\O(K^2)$ terms,  we obtain
\begin{align}
&\frac{\tl j^0(q)}{\pi}=-e  \(2\delta(\Delta q ) + \(({q^x} \k e^{i\f}-i q^y) \frac{K}{2\g k} \delta_{\frac{L}{\b^z}}(\Delta q+ k \beta^z)-({q^x} \k  e^{-i \f}+i q^y) \frac{K}{2\g k} \delta_{\frac{L}{\b^z}}(\Delta q- k \beta^z)\)\),\\
 &\frac{\tl j^x(q)}{\pi}= \frac{e K  }{2\g } \k e^{i \f}\delta_{\frac{L}{\b^z}}(\Delta q+\beta^z k)+\frac{e K}{2\g }  \k e^{-i \f} \delta_{\frac{L}{\b^z}}(\Delta q-\beta^z k),\\
 & \frac{\tl j^y(q)}{\pi}= \frac{e K}{2\g i}\delta_{\frac{L}{\b^z}}(\Delta q+\beta^z k)-\frac{e K}{2\g i}\delta_{\frac{L}{\b^z}}(\Delta q-\beta^z k), \\
 &\frac{\tl j^z(q)}{\pi}=-e  \beta^z \(2\delta(\Delta q)+\(({q^x} \k e^{i\f}-i q^y) \frac{K}{2\g k} \delta_{\frac{L}{\b^z}}(\Delta q+ k \beta^z)-({q^x} \k  e^{-i \f}+i q^y) \frac{K}{2\g k} \delta_{\frac{L}{\b^z}}(\Delta q- k \beta^z)\)\).
\end{align}
Here,
\beq
\Delta{q}\equiv q^0-\beta^z q^z
\eeq
and $\pi\delta_X(q)\equiv \int_{0}^{X}{ d t e^{i q t}}$ is the `delta-function' with a finite width.
This delta-function appears in the coefficient of $K$  because it is related to the length of the magnetic field, while the usual delta-functions in $\tl j^0, \tl j^z$ do not vanish when $K\to 0$. 
The $K$-dependent terms in $\tl j^0, \tl j^z$ originate from the Taylor expansion of the exponent by $K$, similar to the approach used in perturbation theory in QFT. A more precise calculation, which will predict not only the fundamental radiation but aslo higher order harmonic radiation in the context of conventional undulator radiation, can be performed by using the precise solution of the electron motion and including higher-order terms of the expansion.\footnote{An even more precise estiamtion can be made by taking electron field into account in the Lagrangian and perfomring the path integrals including the electron and photon loops in addition to the background current and magnetic field. }

Then, \Eq{amp} becomes
\begin{align}
\sqrt{ 2 w_\g}\frac{\langle{\g,\e,k_\g|0_j\rangle}}{\pi}\simeq  
i\frac{e K}{2 k \g} \(k( \k e^{-i\f}\e^x + i\e^y ) +\e^z (\k e^{-i\f}k_{\g}^x +ik_{\g}^y)\beta^z\) \delta_{\frac{L}{\b^z}}(\D k_\g -k \beta^z).
\laq{ampsimp}
\end{align}
Here, $\cos \theta = \frac{k_{\g}^{z}}{w_\g}$
and $
\Delta{k_\g}=w_\g(1-\beta^z  \cos(\theta)).
$
In deriving the amplitude, we have neglected terms proportional to $\delta(\D k_\g), \delta_{\frac{L}{\b^z}}(\D k_\g +k\beta^z)$, which are suppressed because their arguments cannot be zero. As conventional, we take $\e^0=0$, and the two polarization vectors of photons are given by $\vec{\e}_\mp=\frac{1}{\sqrt{2}}\(-\cos\h \cos \tl \f, -\cos\h \sin \tl \f, \sin\theta\)\pm \frac{i}{\sqrt{2}} \(\sin\tl \f, -\cos\tl \f,0\)$
 with $\tl \f$ satisfying $k_{\g}^{x}= k_\g \sin \theta \cos \tl\f, k_{\g}^{y}= k_\g \sin \theta \sin \tl\f,  k_{\g}^z=k_\g\cos \theta $.\footnote{This is  the definition of $\e_\mp$ rather than $\e_\pm$, as we consider the outgoing photon. }

The total energy of the photon to be produced can be estimated as 
\begin{align}
J =\int \frac{d k_\g  k_\g^2 d\Omega}{(2\pi)^3} w_\g \sum_{\e} \ab{\langle{\g,\e, \vec k_\g|0_j\rangle}}^2.
\end{align}
To avoid writing a lengthy form and check the consistency with \Eq{photonflux}, let us take $\k=1,\f=0$ again. 
Then we get
\begin{align}\laq{J}
J&\approx \int {d \Omega} {\frac{L}{\b^z}} \frac{e^2 K^2 k^2\beta_z^2}{16\pi^2\g^2}\frac{1}{(1-\cos \theta \beta^z)^3}
\(1- \frac{\g^{-2}(2-\cos\theta)\sin^2\theta+\O(\g^{-4})}{2 (1- \cos \theta \beta^z)^2}\),\\ 
w_\g &=\frac{\b^z k}{1-\cos\theta\beta^z} 
\laq{wg}
\end{align}
Here, we used 
$\int_{-\infty}^{\infty}{d k\ab{\pi\delta_L(f(k))}^2}=2\pi L /|f'|$. The second equation is from the `delta-function'. Note again that this delta-function has a finite width. 
 Thus the resulting photon spectrum will have the corresponding width $\D k_\g/k_\g\sim  \pi \beta^z /(L k)$ around \Eq{wg}.
By dividing the integrand by the electron propagating time, $L/\b^z$, and approximating $\b\simeq \beta^z$, we get $d P/d\Omega$. This agrees with \Eq{photonflux} very well. 
The agreement up to the correction term of $\g^{-2} \sin^2\theta/(2(1-\cos \beta^z)^2)$ in \Eq{photonflux} is obtained when we expand the exponent by $-iq^x r^{x}-i{q}^{y} {r}^{y}$ in \Eq{tljm}. This means that those terms from the oscillating exponential in $j^0,j^z$ are also relevant to the synchrotron radiation  
because, in the Maxwell equation estimations, we did not use the Coulomb potential. 
In this case, the photons are helical and predominantly composed with the polarization of $\e_+$ with $|\theta|\ll 1/\g$, as can be found in \Eq{ampsimp} at the amplitude level.

By performing the integral over solid angles, 
we obtain the total injected energy per unit time, 
\beq
W\approx  \frac{(1+\k^2)e^2 k^2 K^2}{12\pi \g^2(-1+\beta_z^2)^2} \approx  \frac{(1+\k^2) }{(1+\frac{1}{2}K^2 (1+\k^2))^2}\frac{e^2 k^2 K^2 \g^2}{12\pi }.
\eeq
We have recovered the $\k,\phi$ dependence and show the result up to the leading order of the $K, 1/\g$ expansions. 
The total number of photons emitted per unit time can be similarly obtained by dividing the integrand of \Eq{J} by $w_\g$ and performing the integral
\beq
\laq{ngamma}
\dot{n}_\g\approx   \frac{(1+\k^2)e^2 k K^2}{12\pi \g^2(-1+\beta_z^2)} \approx  \frac{1+\k^2}{1+\frac{1}{2}K^2 (1+\k^2)}\frac{e^2 k K^2 }{12\pi }.
\eeq
We note that the total number of photons does not exhibit the enhancement with  $\g$.

\section{ALPs from Undulators}
\lac{ALP}
Now, we are ready to discuss the main topic, ALP production. Here we consider the case of a single ALP production for simplicity, while our analysis can be easily extended to study many ALP productions.\footnote{When the ALP mass differences are larger than the inverse of the distance between a detector and the undulator, the production rate can be estimated individually. Namely we calculate \Eq{ntot} for each ALP and then sum them up.}
Given an external magnetic field, $\vec{B}=\vec{B}_{\rm ext}$, 
it is often discussed that a photon wave can be converted into an ALP, $\phi$, with the mass, $m_\f$, 
due to the ALP interaction Lagrangian
\beq
\laq{coupling}
{\cal L}_{\rm int}=-g_{\f\g\g}  \f \vec{E} \cdot \vec{B}.
\eeq 
Again, we consider the electrons propagating in the $z$-direction from $z=-\infty $ to $\infty$ following the equation of motion under the influence of the external magnetic field. 

The conversion of asymptotic state photons to ALPs is commonly studied in the presence of periodic or constant magnetic fields~\cite{Abel:2006qt,Arias:2010bh,Bahre:2013ywa,Ortiz:2020tgs,Hoof:2024gfk}.
The axion-photon conversion rate scales with $(g_{\f\g\g} B_0 L)^2$ in the case of the uniform magnetic field ($k\to 0$) or with an oscillating magnetic field when the ALP mass is tuned for resonance.
The production of massive ALPs from massless photons can be easily understood as a result of the violation of momentum conservation due to the presence of a stationary magnetic field in a finite volume.\footnote{If the external field is time-dependent, energy conservation may also be violated. In some experimental setups, by searching for line photons converted from ALPs, one needs to be careful of the finite width of the `delta-function' for energy conservation.}

Here, we develop formalisms for calculating the ALP production from the first principle and estimate the number and spectra of the ALPs from undulators. 

\subsection{Field-Theoretic Estimation of ALP Production}
\lac{ALPprod}
Let us first provide some basic formalism, which works well, especially for light ALP production. This formalism will be changed to apply to heavy ALP production with resonance effect in \Sec{reso}.

Similar to our estimate of the photon production based on QFT, we estimate the ALP production 
by treating the electron current $j^\m$ and the external magnetic field $B_{\rm ext}$ as the background fields. 
The corresponding diagram is shown in the right panel of Fig. \ref{fig:Feynman}.
The transition amplitude can be estimated as
\beq\laq{genericform}
\langle \f, k_\f |0_{jB}\rangle =\frac{g_{\f \g\g}}{\sqrt{2w_\f}}\int d^4x d^4y e^{i(w_\f t-\vec{k}_\f \vec{x})}\sum_{l=x,y,z} B_{\rm ext}^l(x)\( \partial_0 \Delta_{l\m}(x,y)- \partial_l{\Delta_{0\m}(x,y) }\) j^\m(y).
\eeq
Here, $\D_{\m\n}(x,y)=\int d^4q e^{-i q \cdot (x-y)}\frac{-i \eta_{\m\n}}{q^2+i\e}$ represents the photon propagator, and we adopt Feynman gauge for ease of calculation; 
and $(w_\f,\vec{k}_\f)$ is the four-momentum for the ALP, satisfying $w_\f=\sqrt{k_\f^2+m_\f^2}.$ By performing the $x^\m, y^\m$ and $q^{0,x,y}$ integrals, and given that $\vec B_{\rm ext}$ does not vary in $x,y$
directions, 
 we obtain
\begin{align}
\sqrt{w_\f}\langle \f, k_\f |0_{jB}\rangle &=\int{\frac{dq^z}{(2\pi)}\frac{\pi}{2}\frac{g_{\f\g\g} B_0}{q^2+i\e}\left( \delta_L(-k_{\f}^z+q^z+k) \(q_0 \vec \e_B \cdot \non\vec{\tl{j}}(q)+\vec{q} \cdot \vec{\e}_B\tl{j}_0(q)\)\right.}\\
&\hspace{33mm}{+\left. \delta_L(-k_{\f}^z+q^z-k) \(q_0 \vec \e_B^* \cdot \vec{\tl{j}}(q)+\vec{q} \cdot \vec{\e}_B^*\tl{j}_0(q)\)\)}.\laq{ampgene}
\end{align}
Here, the delta-functions that are integrated out set $q_0=w_\f, q_{x}=k_{\f,x}, q_y=k_{\f,y}.$ 
We defined $\vec{\e}_B=\(1/\sqrt{2},-i\k e^{i\f}/\sqrt{2},0\),$ satisfying $\vec{B}_{\rm exp}=\sqrt{2}\Re[B_0\vec\e_B e^{ikz}]$.

In the following,  we approximate that $\d_L,\AND \d_{L/\b^z}$ are the usual delta-functions as before and drop all of the delta-functions whose argument is not zero, assuming $k_{\f}^z\gg k$. 

\paragraph{Momentum-Polar-Angle Relation of Undulator ALPs} Adopting the approximation, we obtain
\begin{align}
&\frac{\langle \f, k_\f |0_{jB}\rangle}{g_{\f\g\g} B_0 L} \simeq 
\laq{amp1} \frac{e (k_{\f}^x+i\kappa   k_{\f}^y  \exp (-i \phi))}{2  \sqrt{2w_\f} \left(\g^{-2} w_\f^2/\beta^z+k_{\f,x}^2+k_{\f,y}^2\right)} \d_{\frac{L}{\b^z}}(w_{\f}-\beta^z  (k+k_{\f}^z))\\
\laq{amp2}&-\frac{eK e^{-2 i \phi} \left(\kappa  k_{\f}^x+i k_{\f}^y e^{i \phi}\right) \left(k_{\f}^x e^{i \phi}+i \kappa  k_{\f}^{y}\right)}{8 \gamma  k \sqrt{2w_\f} \left(-w_\f^2+(w_\f-k \beta^z)^2\beta_z^{-2}+k_{\f,x}^2+k_{\f,y}^2\right)}\delta_{\frac{L}{\b^z}}(w_{\f}-\beta^z  (2 k+k_{\f}^z)),
\end{align}
where we also used $\d_L(X/\beta^z )=\beta^z \d_{L/\b^z}(X)$.
From the form of the equation, it is evident that the amplitude vanishes when $k_{\f,x}=k_{\f,y}=0$, indicating that ALP production is suppressed in the forward direction. The ALP momenta have the form
\beq\laq{ana}
k_{\f}=
\frac{\sqrt{k^2 \beta _z^2+m_{\phi }^2 \left(x^2 \beta _z^2-1\right)}+k x \beta _z^2}{1-x^2 \beta _z^2},~~~~
\frac{\sqrt{4 k^2 \beta _z^2+m_{\phi }^2 \left(x^2 \beta _z^2-1\right)}+2 k x \beta _z^2}{1-x^2 \beta _z^2}
\eeq
to set the arguments of the first and second delta-functions, respectively, zero. They give relations between the polar angular, $\theta,$ and $k_\f$. 
Since the delta-function $\d_L(x)$ is not exact and has a finite width, given by $\D x=\pi/L$, we will have the spectra with the width of 
$
\frac{\D k_{\f}}{k_\f}\sim  \frac{\pi }{L k},
$
 around \eq{ana}. We can also have higher momentum modes if we expand $K$ terms in the exponent of the electron current up to higher orders.

\begin{figure}[t!]
    \begin{center}  
    \includegraphics[width=145mm]{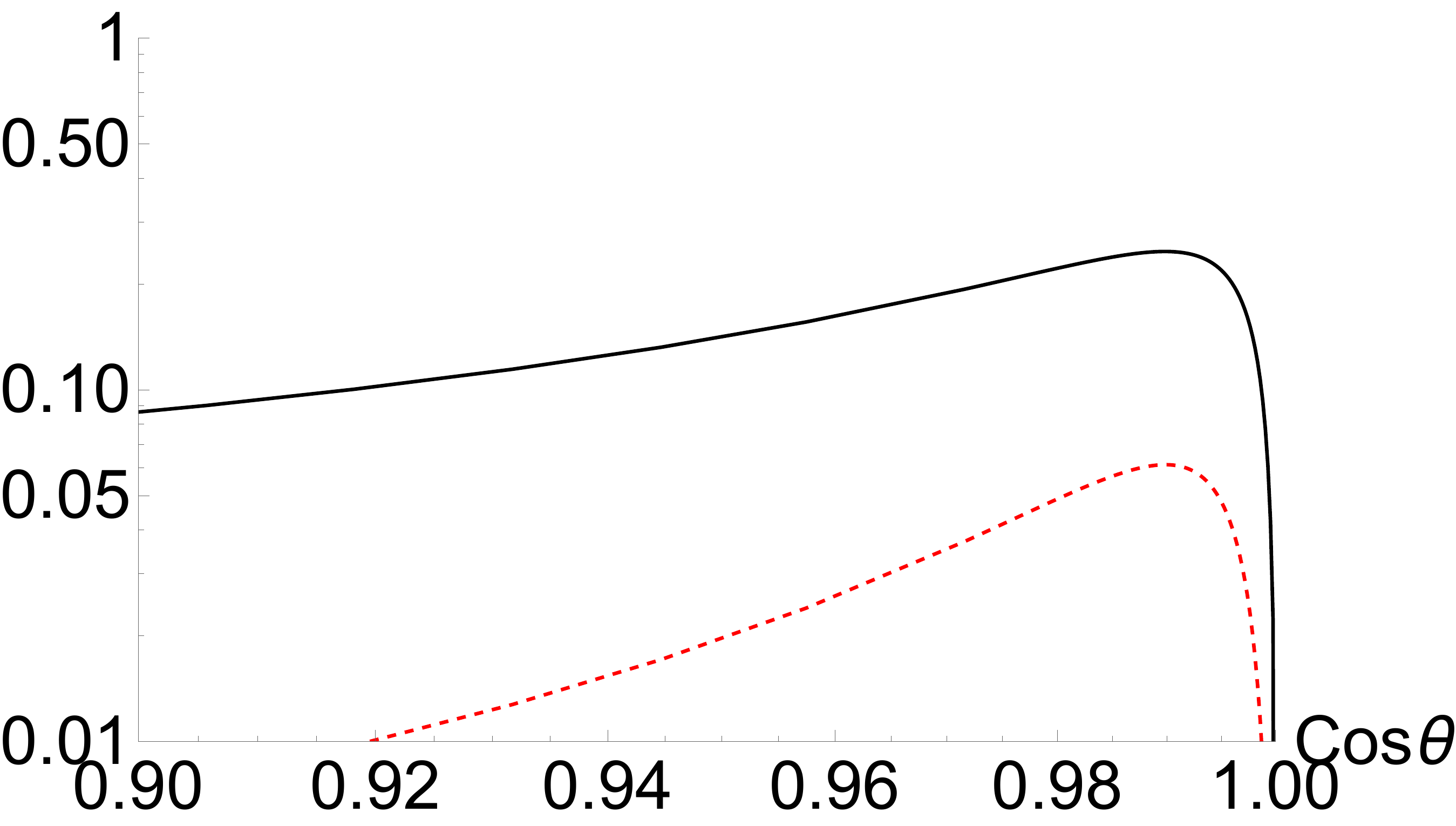}
    \end{center}
    \caption{ The angular dependence of the first term (Black solid line) and the second term (Red dashed line) of the integrand of \Eq{numberdis2}.  
    For simplicity, we take $\beta^z=0.99, e^2 (g_{\f \g\g}B_0)^2 L/k =1$,  $K=1$, $\k=1$, and $\f=0$. The dependence on them can be easily recovered from \Eq{numberdis2}. 
 }
    \label{fig:dist}
    \end{figure}

\paragraph{Spectra of Undulator ALPs} The resulting total number of the ALP can be obtained from 
\beq
 \laq{numberdis}
n_\f=  \int \frac{d^3 \vec{k}_\f}{(2\pi)^3}|\langle \f, k_\f |0_{jB}\rangle|^2.
\eeq
For simplicity, assuming $m_\f=0$, we can then derive the analytic formula of the produced ALP number: \beq \laq{numberdis2}
n_\f\simeq \frac{{g_{\f\g\g}^2 B^2_0 e^2}L}{32\pi^2 k}\int dx \frac{f_1(\beta^z)^2 \left(1-x^2\right)}{ \left(\beta_z ^2 x^2-1\right)^2}+\frac{2 f_2 (\beta^z) ^4 K^2 \left(x^2-1\right)^2}{(\beta^z  x-1)^4 (\gamma +3 \beta^z  \gamma  x)^2}
 \eeq
where $x$ represents $\cos \theta$, and $f_1=\pi (1+\k^2),~f_2= \frac{\pi}{4}\( \kappa ^4+2 \kappa ^2 \cos (2 \phi)+4 \kappa ^2+1\)$ satisfying $f_{1,2}=2\pi$ for $\k=1.$ 
 The first term in the integrand comes from \Eq{amp1}, while the second term comes from \Eq{amp2}.\footnote{We do not consider the divergence at $x=- 1/(3\beta)$ for the second term because the assumptions setting various delta-functions to zero do not hold in this case.
Numerically, this divergence is not found. However, we verified that the dominant production of $\phi$ at $x=-1$ due to the first term is physical. Interestingly, it corresponds to suppressed ALP momentum $k_\f \sim k/2.$
This behavior will be used to discuss ALP searches in the future. 
  } 
 We consider the ALP production to be in the forward direction. From the both integrands, we see that ALPs are mostly produced in the direction of
 \beq
\sin\theta \simeq \frac{\sqrt{1-\beta_z^2}}{\beta^z}\simeq \frac{\sqrt{1+\frac{1+\k^2}{2}K^2}}{\g},
 \eeq
 at which we have \beq k_\f\simeq \frac{\g^2 k}{1+\frac{1+\k^2}{2}K^2},~~ \frac{2\g^2 k}{1+\frac{1+\k^2}{2}K^2}\eeq 
 for the $f_1$ and $f_2$ terms, respectively. 
The polar angular dependence of the first term and the second term is shown in Fig.~\ref{fig:dist}, where we take $\beta^z=0.99$, $K=1,\k=1,\AND \f=0$. One can see that the ALP is dominantly produced at $\theta \sim 1/\g$, but is suppressed when $\theta \ll 1/\g.$ The $K^2$ term is subdominant even when $K=1$ with the not too large $\g.$

We also mention that the dependence of the azimuthal  angular $\tl{\f}$ satisfying $\{k_{\f}^x,k_{\f}^y\}=\{k_\f\sin\theta\cos \tl \f,k_\f\sin\theta\sin \tl \f\})$ can be easily recovered from our estimate. 
From the integrand of \Eq{numberdis}, the Coulomb contribution term proportional to $f_1$
has the dependence of 
\beq
\laq{angdep}
\partial_{\tl{\f},f_1} n_{\f}\propto \kappa  \sin (\tl \phi) (\kappa  \sin (\tl \phi)+2 \cos (\tl \phi) \sin (\phi))+\cos ^2(\tl \phi )\eeq
The contribution proportional to $f_2$ has
\begin{align}
\laq{angdep2}
\partial_{\tl{\f},f_2} n_{\f}\propto &\(1+4 \kappa ^2+\kappa ^4\)+4 \kappa  \left(\left(\kappa ^2-1\right) \sin (4 \tl \phi ) \sin (\f)+\kappa  \sin^2(2 \tl \phi ) \cos (2 \f) \right)\non\\
&-\left(\kappa ^4-4 \kappa ^2+1\right) \cos (4 \tl \phi ).
\end{align}
This correspondence satisfies throughout this paper.
For instance, a linear undulator with $\k=0$ has the preferred direction of $\tl{\f}=0,\pi$ for $f_1$ term and $\tl \f=\pi/4,3\pi/4,5\pi/4, 7\pi/4$ for $f_2$ term. The $f_2$ term contribution is shown in Fig.\,\ref{fig:concept}. On the other hand for $\k=1, \f=0$ we have $\tl \f$ independent spectrum (c.f. Fig.\ref{fig:contour2}).

\paragraph{Total flux of light Undulator ALPs} The total ALP number flux in the forward direction can be estimated by integrating over $x$ from $1/3$ to 1 and dividing the result by $L/\b^z$. Here, the lower limit will not affect our result in the leading order of $1/\g$ and $m_\f$ expansion.   
At this leading order, this turns out to be
\beq
\laq{nf}
\boxed{\dot n_\f\simeq \frac{e^2 \(g_{\f \g\g}B_0\)^2 }{64\pi^2 k}\left\{\(2 \log(\frac{2\gamma}{\sqrt{1+\frac{1+\k^2}{2}K^2}})-1\)f_1+\frac{2}{3}\frac{K^2}{1+\frac{1+\k^2}{2}K^2} f_2\right\}.}
\eeq
Due to the logarithmic enhancement, $f_1$ term is always dominant with $m_\f\ll k$. A large $K$ does not help enhancing the second term because $\lim_{K\to \infty}\frac{K^2}{1+\frac{1+\k^2}{2}K^2}\to\frac{2}{1+\k^2}.$ We conclude that with $m_\f\ll k$, we have dominant ALP production from the Coulomb term.

From \Eq{ngamma},  we can estimate the effective axion-photon conversion rate, $\eta$, by neglecting the $K$-dependent term in \Eq{nf},
\beq
\laq{eta}
\eta\equiv \frac{\dot n_\f}{\dot n_\g}\approx \frac{3\(g_{\f\g\g}B_0\)^2}{16K^2 k^2}\(2 \log (2\gamma)-1\)=\O(0.1)\frac{\(g_{\f\g\g}m_e\)^2}{e^2}
\eeq
which, interestingly, does not depend on $\k$ and $\phi$ at the leading order.
This conversion rate is equivalent to placing $1$~Tesla magnetic fields over approximately $0.002$~m. 
In this sense, the number of produced ALPs is lower than in original LSW experiments, which typically use Tesla-level magnetic fields over a few meters. 

When increasing the ALP mass to $m_\f\gtrsim k$, we have enhanced ALP production. 
We find that the resulting ALP number diverges in the contribution proportional to the squared of \Eq{amp2}
when integrating over  $ 0 \ll \cos\theta<1$.  The divergence is due to the approximation where we treat the delta functions as exact, as we will see in \Sec{reso}.

Regarding to the subdominant $\O(K^2)$ term, we also comment on a contribution from the effect of the width of the `delta-function'. 
As we have mentioned, the delta-function $\d_X (y)$ is not exact (and it is even a complex function in our definition).  For a value of $y\gg \pi/X$, we have $\pi\delta_X(y)\sim \O(\frac{1}{y})$. Usually, the contribution is suppressed; however, one possibly relevant enhanced contribution is from the on-shell photon production corresponding to the pole of the propagator of the diagram in the right panel of Fig.\,\ref{fig:Feynman}, which will be also discussed in detail in \Sec{reso}. 
The amplitude after the $q_z$ integration over the pole is the sum of the products of $\delta_{L}\delta_{L/\beta^z}$. We can check that only one of the two delta-functions in any product can have its argument zero from the kinematics with $m_\f\ll k$. Then, the corresponding ALP production rate scales as $\propto \O(g_{\f\g\g}^2 B_0^2 \frac{e^2}{y} K^2)$.
  This, at most, gives a contribution as large as subdominant $\O(K^2)$ term. 
  Thus, we neglect it in the ALP number production.\footnote{However, the preferred direction for the produced ALP is different from the Coulomb one, and, in this sense, it is not completely irrelevant with $K=\O(1)$. This contribution can also be checked in the numerical simulation formulated in \Sec{nume}.} 
  As we will see, if $m_\f \gtrsim k$, both arguments of the products of the two delta-functions can be zero kinematically, with the photon on-shell conditions, which can give an enhancement of the ALP production.

\subsection{Field-Theoretic Estimation of ALP Production withPhoton Resonance}
\lac{reso}

A parametrically heavy ALP gives a divergent number density from the $K^2$ term in the previous estimation. This is because, at certain $k_\f$, $\cos\theta$ and $q^z$, not only both arguments of the products of the two delta-functions in the \Eq{ampgene} but also the numerator of the photon propagator (see the right panel of Fig.\,\ref{fig:Feynman}) can be zero kinematically, i.e., a `resonance' exists. 
However, in the previous estimation, we integrate the two `delta-functions'  rather than the true delta-function from the pole of the photon propagator, giving the divergence. Our previous approximation of exact delta-functions fails in this case.

To study the $m_\f \gtrsim k$ regime, let us first integrate real delta-function from the photon pole, $\Im[\frac{1}{q^2+i\e}]= -i \pi \delta(q^2)$
in \Eq{ampgene} when we perform the $q^z$ integral. We get 
\begin{align}
\sqrt{w_\f}\langle \f, k_\f |0_{jB}\rangle = & i \frac{ eK  e^{-2 i \phi} 
\left(\kappa  k_{\f}^x+i k_{\f}^y e^{i \phi}\right) \left(k_{\f}^x e^{i \phi}+i \kappa  k_{\f}^{y}\right) 
}{16 \sqrt{2} \gamma  k w_\f^{3/2}} \pi \d_L(F_1)\times \pi \d_{\frac{L}{\beta^z}}( F_2)
\end{align}
with 
\beq
F_1=\sqrt{k_\f^2 x^2+m_\f^2}-k_\f x-k,~~~ F_2=\sqrt{k_\f^2+m_\f^2}-\beta^z  \sqrt{k_\f^2 x^2+m_\f^2}-k\beta^z
\eeq
We note that $F_1=0$  cannot be satisfied if $m_\f \ll k$ with $k_\f\gg k.$ Thus, we cannot have efficient production of the ALP via this `photon resonance' as discussed at the end of the \Sec{ALPprod}. When $m_\f \gtrsim k,$ we can also calculate the produced number of ALPs by performing the integral \eq{numberdis}
by approximating that the integral of the product of four delta-functions with respect to $k_\f$ and $x$ gives $\frac{(2\pi L)^2/\beta^z}{|{\rm det}\partial_{k_\f,x} F_i|}$. 
By performing the expansion of $k$ and $\g^{-1}$, we arrive at the simple formulas 
\beq
\boxed{
n_\f^{\rm reso}\simeq g_{\f\g\g}^2 B_0^2 L^2 e^2 \frac{K^2}{512 \pi } \frac{m_\f^2}{ k^2 \g^2} f_2}
\eeq
in which the ALPs are mostly produced with 
\beq
k_\f\simeq \frac{m_\f^2}{2k}, ~~~~x\simeq 1-\frac{2 k^2}{m^2_\f}
+\frac{1+\frac{1+\k^2}{2}K^2}{2\g^2}.
\laq{resopeak}
\eeq
From $x\leq 1$ we obtain $\frac{2k}{m_\f}\gtrsim \sqrt{1+\frac{1+\k^2}{2}K^2}\g^{-1}.$
Then, we get the condition  to have the resonance 
\beq
\laq{condm}
k^2\lesssim m_\f^2\lesssim \frac{-k^2 + 3 k^2 \beta^z}{1 - \beta^z},
\eeq
where we solved $F_1=0,F_2=0, x=1$ for the upper bound. 
At the upper bound, $x=1$, a $m_\f^4 k^4 \O(\g^{-4})$ term becomes comparable, canceling the contribution we have shown, and $n_\f$ goes to zero. This behavior is suggested from the symmetry in the helical undulator limit (see \Sec{symmetry}).\footnote{The symmetry argument also explains the non-helical case because the  $\k,\f$ dependent terms factorize.} 
Since the delta-functions are not exact ones, not only $k_\f$ but also $x$ has a narrow width $\D k_\f/ k_\f, \D \theta/\theta  \sim \pi/(L k).$\footnote{This is interesting because it implies for large $L k/\pi$, the ALP can be searched for with a not too large detector at a faraway place.}

Satisfying \Eq{condm}, the effective axion-photon conversion rate (see \Eq{eta} for the definition) can be estimated as 
\beq
\eta^{\rm reso}=g_{\f\g\g}^2 B_0^2 \frac{3 \pi L}{128 k}f_2 f_1^{-1} \frac{m_\f^2}{\g^2 k^2}
\eeq
For relatively large $K$ and large $L$, this gives a dominant contribution compared to the previous production via off-shell photon. 
The number of ALP is maximized when $m_\f \sim 1/\sqrt{1-\b_z^2} k\sim \g k$. 
For instance, a typical single undulator in  SPring-8 or NanoTerasu, which have $\O(1)$Tesla magnetic field and length $\O(1-10)$m, with $k=\O(10)$cm, 
is equivalent to the LSW experiment with placing an approximately $1$ Tesla magnetic field over $\O(0.1)$m, which is not very small given that we can perform the experiment during the whole operation time of the facility, say $\O(1)$ years. In addition, this contribution is interesting because it can produce heavier ALPs compared with the conventional LSW experiments with a spatially constant magnetic field.

\begin{figure}[t!]
    \begin{center}  
    \includegraphics[width=145mm]{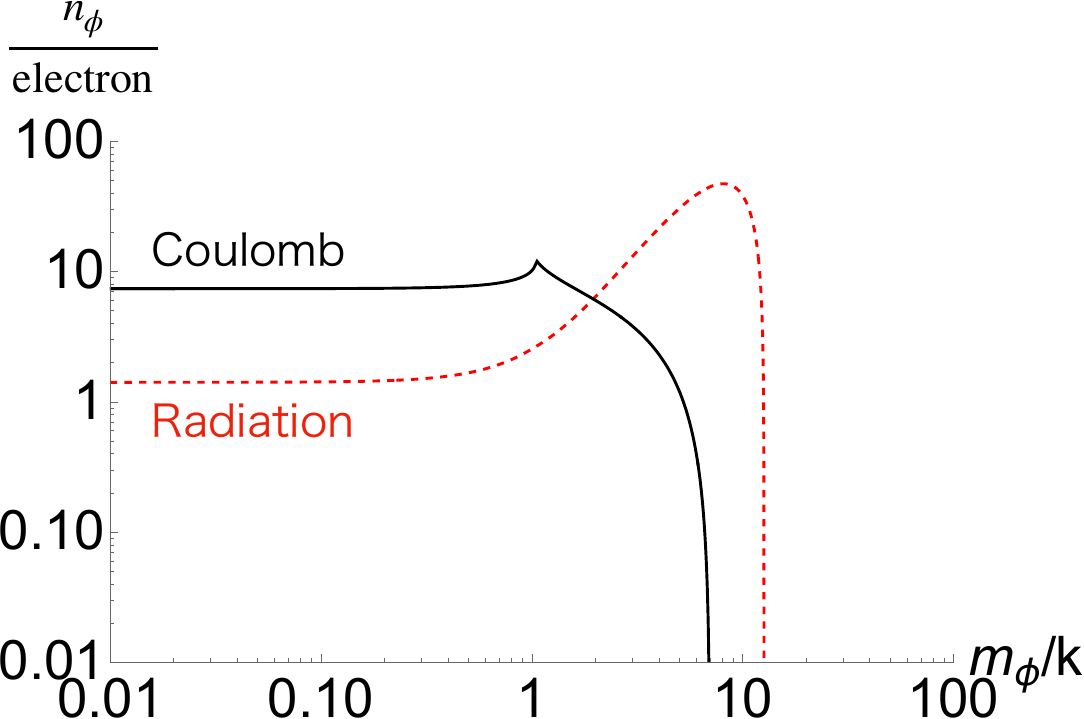}
    \end{center}
    \caption{The produced ALP number when an electron passes through the undulator by varying the ALP mass.  
    The $K^0$ order term is shown in the Black solid line, and the $K^2$ term, including the resonance contribution, is shown by the red dashed line. We take 
     $\gamma=10, e^2 (g_{\f \g\g}B_0)^2/k =1$,   $\k=1,\f=0$, $K=1$ and $L=33 \times 2\pi/k$.
 }
    \label{fig:mdep}
    \end{figure}

The total contribution of the ALP production can be obtained with the na\"{i}ve summation of the two contributions,
\beq\laq{ntot}
\dot{n}^{\rm tot}_\f[m_\f]=  n_{\f}+n_{\f}^{\rm reso}\eeq
In the $x$ integral of $n_\f$, we used the boundary condition of $x=[\max(1/3,\sqrt{1-\beta_z^2 k^2/m^2}/\beta^z),1]$ to remove the singularity set by the resonance. The $\O(K^0)$ term and $\O(K^2)$ terms are shown in Fig.\ref{fig:mdep}.


\section{Conclusions and discussion }

In this paper, we have developed a formalism for estimating the ALP production in the ALP, photon, charged current system from a first principle QFT calculation to show that ALPs can be produced in any undulator with a standard design.
Light ALP was found to be dominantly produced from the Coulomb potential of the electron propagating in the undulator and the installed oscillating magnetic fields for the electron to undulate. For relatively heavy ALP, a resonance effect can become important, enhancing the production rate significantly.  
 These contributions are unavoidable by tuning the undulator parameters if the ALP exists in the relevant mass range. 
Importantly, the preferred direction of the ALP production is always misaligned with the preferred direction of the photon production. The detailed favored direction and the spectra of the ALPs depend on the property of the undulator, and the formulas for them have been derived.

This conclusions implies that if we put a detector, e.g., the LSW like detector for the axion-photon coupling, 
or even more faraway photon detectors after the geomagnetic field\footnote{For instance, if we employ an undulator with electron energy of $\O(100)\GEV$, which is discused in the future collider concepts, the off-axis angle of ALP production is $(1/\g)\sim 10^{-6}$. We need the detector of size $10^{-2} {\rm m} \(100{\rm km}/{\rm distance}\)$ placed at $d$ to cover the preferred direction of the ALP production. If we focus on the resonant regime, a smaller detector may be possible since the ALP is produced in a suppressed polar angle regime satisfying $\Delta \theta/\theta \lesssim \pi/(k L)$. An Earth observation satellite with a photon detector may be an effective detector. 
}, on the direction, we just wait for the detection of ALP while performing the experiments of the facility that initially aimed.\footnote{Some ALPs may not only have photon couplings but also couplings to the other standard model particles, and one can examine the product of the photon coupling and the other coupling as well by placing a certain detector. The most economical one may be the direct detection experiments of WIMP by considering the ALP absorption by heavy atoms and checking the electron recoils. In this case, we also need to study if the electron coupling is important for producing the ALP in the undulator.} The preferred direction of ALP production may even be useful in reducing background noise. 

Given our analytical formulas, it is straightforward to estimate the sensitivity reach of probing the undulator ALPs. 
Let us consider a facility with an undulator by the operating power of $P$ over a year. 
Then the produced photon number over the year should be 
\beq
\dot{n}_\g \approx \eta_w \frac{P}{w_\g}\approx  10^{27}/{\rm yr} ~ \frac{\eta_w P}{\rm 10 kW} \frac{1\KEV}{w_\g}.
\eeq
Here $\eta_w$ is the efficiency parameter denoting the fraction of the energy to be converted to the photon energy. Then we get \beq 
\dot n_\f= \dot{n}_\g (\eta+\eta^{\rm reso}).\eeq

\begin{figure}[t!]
    \begin{center}  
    \includegraphics[width=145mm]{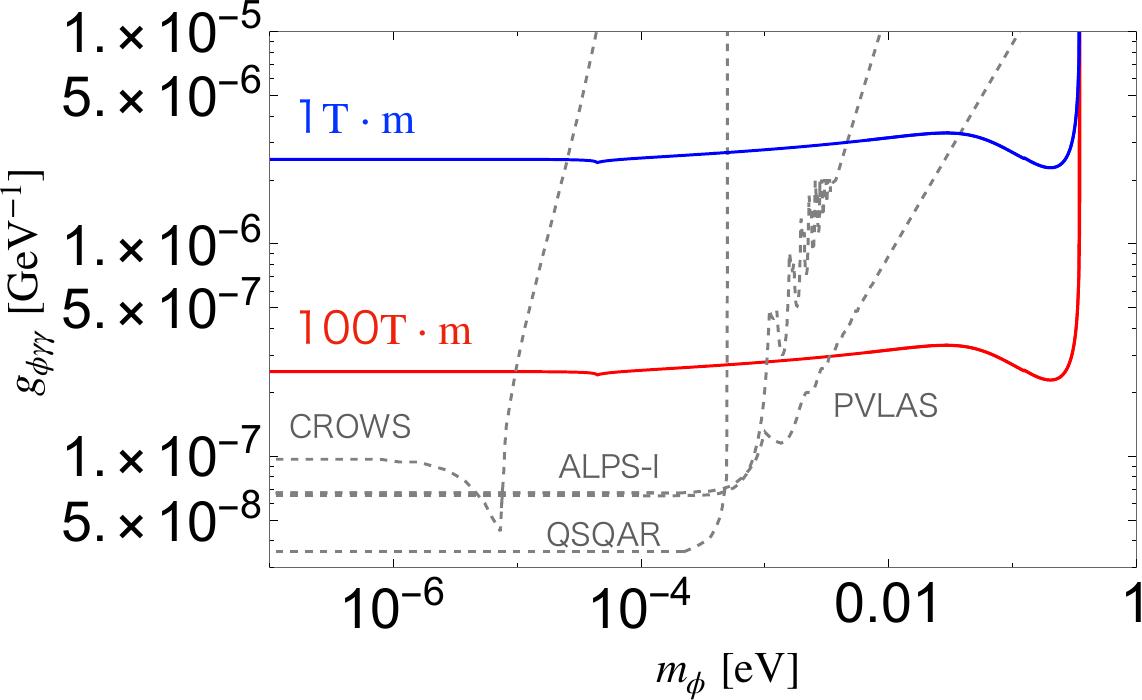}
    \end{center}
    \caption{
    An example plot of the sensitivity reach of the undulator ALP search by placing a detector in $m_\f, g_{\f \g\g}$ plane. 
We assume the detector efficiency $\sim g_{\f\g\g}^2 0.24 \GEV^2 $(blue solid line), and $ g_{\f\g\g}^2 2400 \GEV^2 $ (red solid line),  $P\eta_w={\rm 15 kW}, w_\g=2 k \g^2, \g=3\GEV/m_e, 2\pi/k=30$cm, $L=33 \times  2\pi/k$, and a run duration of 3 years.
Some exsiting experimental results from LSW experiments are also shown for comparison.  
 }
    \label{fig:5}
    \end{figure}

    An example plot of the reach of ALP produced from the undulator (Red solid line) in $m_\f, g_{\f \g\g}$ plane is shown in Fig.\ref{fig:5} with
$P\eta_w =15 {\rm kW}, w_\g=2 k \g^2, \g=3\GEV/m_e, 2\pi/k=30$cm, $L=33 \times  2\pi/k$. 
Here, we assumed detectors with efficiencies $D\sim g_{\f\g\g}^2 2400 \GEV^2$ (red solid line) and $D\sim g_{\f\g\g}^2 0.24 \GEV^2$ (blue solid line), namely the ALP events can be estimated with $N=\dot{n}_\f \times D \D t_{\rm exp}$ with $\D t_{\rm exp}$ being the experimental time.
Although we do not specify the property of a detector, this efficiency corresponds to the axion-photon conversion of the product of the magnetic field and the effective length $100\rm Tesla \cdot m$ and $1\rm Tesla \cdot m$ magnetic field.\footnote{Strictly speaking, for heavy ALP, we need an undulator-like detector for having efficient axion-photon conversion in the case of the axion-photon coupling. $1\rm Tesla \cdot m$ is satisfied for an undulator magnet. }  
We assumed there is no background, and we set the 2$\s$ reach for the ALP production (the red solid line), i.e., $N=4$ events. The experimental results from LSW experiments \cite{Ehret:2010mh,Betz:2013dza,OSQAR:2015qdv, Halliday:2024lca} (and PVLAS experiment for the vacuum magnetic birefringenc~\cite{DellaValle:2015xxa}) are also shown in gray dashed lines for comparison. We can see that the reach can be well beyond the present ones at a high mass regime, although the light mass region is typically weaker than the previous LSW experimental searches.\footnote{Our sensitivity reach is suddenly suppressed at a mass of $\O(0.1)\EV$, in contrast to other limits. This occurs because we used an analytic formula for the dominant contribution, approximating the        `delta-function' as exact. Removing this approximation would result in a slower rise in the curve, which, however, depends on the boundaries of the magnetic fields.}

We also comment that the SPring-8 ALP search based on the conventional LSW by using novel pulsed-magnet system~\cite{Inada:2016jzh}, which is out of this figure, can also probe the heavy mass region beyond the other existing LSW experiments. 
Our sensitivity reach is more stringent than all the previous results in the heavy mass range because we take much longer integration time, which is approximately the order of the operation time of the facility due to the fact that the ALP is produced automatically. 
Notably, we do not need to prepare the initial magnetic field and the wall by ourselves, which saves the costs for the experiment. 
Therefore, placing a detector in the desired direction of the undulators is an economical and efficient approach for the ALP search. \\

Our analysis can be improved in some cases. 
Realistically, one uses an electron beam consisting of a series of electron bunches. Our estimation of the total amount of ALP production can also be used in the case of multiple electrons as long as the ALPs are produced incoherently. However, the precise prediction of the ALP distribution needs to take into account the size of the electron bunches and the energy dispersion, which is similar to the usual estimation of the photon spectrum.  
We also mention that in the case of a free-electron laser, such as SACLA in SPring-8,
a large amount of coherent photons is produced by an undulator with very long $L$, implying that the ALP productions should also be enhanced significantly. However, this needs to be checked carefully from the first principles~\cite{Madey:1971zz}, which will be discussed elsewhere. 

We have assumed a generic standard undulator. 
We expect, depending on the setup, that one may produce even more ALPs. 
For instance, a Wiggler has $K\gg1$. 
In this case, the approximation we have used to derive analytic formulas does not apply and we need to reformulate our discussion. A simple way may be the numerical study by modifying public codes of Synchrotron radiation to include the ALP coupling following the analysis in \Sec{nume}. In fact, a synchrotron radiation facility such as NanoTerasu includes several undulators and magnets, and our philosophy of producing ALPs without providing specialized equipment can apply to the more generic cases.
The sweet spot of the NanoTerasu-like facility will be shown in our forthcoming paper~\cite{Yoshida}, which can have more efficient ALP production than the one estimated in this paper. Thus, the ALP number estimated in this paper should be considered as a minimal value that can be produced in a facility containing undulators. 

Our formalism can be applied to more generic systems, such as ALP productions from astrophysical objects that contain magnetic fields to undulate the electron to produce X- or $/\gamma$-rays. Our first principle calculations may give a precise estimation of the produced ALP spectra and the number. 
Also, the QFT analysis can be straightforwardly extended to study the other particle productions from physics beyond the standard model from undulators, such as dark photons, by considering the relevant couplings.

\section*{Acknowledgments}
WY would like to thank Yukinari Sumino at Tohoku University for the helpful discussions and his QFT lecture, where a similar approach to photon radiation from electron current was studied when WY was in his graduate course.
WY also thanks Akimasa Ishikawa for useful comments.
JY would like to thank Susumu Yamamoto at Tohoku University for the discussions on undulators. 
We also thank Toshio Namba at the University of Tokyo for the fruitful discussion on a related topic. 
This work is supported by JSPS KAKENHI Grant Numbers 20H05851 (W.Y.),  21K20364 (W.Y.),  22K14029 (W.Y.), 22H01215 (W.Y.) and Tokyo Metropolitan University Grant for Young Researchers (Incentive Research Fund).

\appendix

\section{Symmetry Arguments for Photon and ALP Productions in Helical Undulators}\lac{symmetry}
So far, we have shown that ALPs can be produced from an undulator by itself. 
The electron Coulomb potential contributes to the production, which is dominant for light ALP production, and the ALP motion has a preferred direction misaligned from the electron motion. 
These are different from the usual photon radiation discussed in \Sec{1}. 
The difference can be understood due to symmetry in the QFT action when $\f=0,\k=1$, and $L\gg \pi/k$. 

\paragraph{Photon production} 
As we have shown in \Sec{1}, the photons are emitted mostly in the electron beam direction, i.e., the $z$-direction, and it is circularly polarized.

Let us first understand photon production from symmetry of the QFT action with the Lagrangian \eq{Lag}.  In the `vacuum' $\ket{0_J}$, with the background current and with $L\to \infty$, we have a symmetry generated by an operator
\beq\laq{Q}
\hat{Q}\equiv -\hat{H}+\beta^z \hat P^z+\beta^z k \hat{L}^z.
\eeq
with $-\hat H, \hat P^z \AND \hat L^z$ being the generators of time translation ($=-$ Hamiltonian), $z$ translation, and rotation around the $z$-axis, respectively. 
Namely, when we put forward the origin of the time axis by $\a$, the $z$-axis by $\a \beta^z$ and rotate the $x$, $y$ plane by $\a k \beta^z  $, the action does not change according to the solution \eq{sol}. 
Separately, they are not symmetries, e.g., $\hat {P}^z$  is explicitly broken due to the existence of an electron (and also a magnetic field). Let us define that $\ket{0_J}$ has $Q=0$ where we use $Q$ without the hat to denote the eigenvalue of $\hat Q$.
Then, the single photon state from the vacuum must have the same charge $Q=0$ according to the conservation. Since a photon has spin unity, 
for the photon moving so close to the $z$-direction that the orbital angular momentum around $z$-direction is vanishing, 
 we have 
\beq
Q=-w_\g+\beta^z k_{\g}^z\pm \beta^z k.
\eeq
One obtains that only when the helicity is plus rather than minus, $Q=0$ is possible. This is the reason we have helical photons injected in the $z$-direction.
In addition, we obtain the relation for photon energy, 
\beq
w_\g(1-\cos \theta \beta^z) = \beta^z k
\eeq
which agrees with \Eq{wg}.

By using the symmetry argument, we can also understand why the Coulomb potential. 
This is because the electron current at the $K\to 0$, where only Coulomb potential is relevant in \Eq{ele}, no longer oscillates, and thus the action with \Eq{Lag}
is invariant by the accidental symmetry
$
\hat{\tl Q}= -\hat{H} + \beta^z \hat{P}^z ~~(K\to  0).
$
$\hat{\tl Q}$ changes the phase of the magnetic field, but in estimating radiation emitted from accelerating electron, this does not matter (see the left panel of Fig.~\ref{fig:Feynman}). 
Thus, the resulting photon must have 
$
w_\g= \beta^z k_{\g}^z
$
which cannot be satisfied with arbitaray $\theta$, since $|\beta^z|<1$.

\paragraph{ALP production} Now, let us turn to the ALP production. 
With $\f=0,\k=1$, the system with the ALP also has the symmetry of $Q$. 

For the ALP moving close in the $z$-direction with neglected orbital angular momentum, the conservation of $Q$ imposes
\beq
w_\f-\beta^z k_{\f}^{z} =0,
\eeq
since the ALP does not have spin.
This never satisfies, meaning that the ALP cannot be produced close to the $z$-direction. 

On the other hand, by taking account of the orbital angular momentum, $l_{\f}^{z}$, around the $z$-direction we have
\beq 
w_\f-\beta^z k_{\f}^z+\beta^z k l_{\f}^{z}=0.\eeq
Quantum mechanics implies $l_{\f}^z=\pm 1,\pm 2\cdots$. This can be satisfied if $l_{\f}^z>0.$
Indeed, they are the arguments of the delta-functions in \Eqs{amp1} and \eq{amp2}, and agree with the angular dependence in the coefficient, e.g., $k_{\f}^x+ i k_{\f}^y\propto e^{i\tl\f}, (k_{\f}^x+ i k_{\f}^y)^2\propto e^{2i\tl\f}$, respectively. 

The reason why the Coulomb potential contributes in the ALP production is as follows. 
The action with \Eq{coupling} does not have the accidental $\hat{\tl Q}$ symmetry even at the $K\to 0$ limit. This is because the translation with $\hat{\tl Q}$ would change the phase of the magnetic field, which couples to the ALP. 
Therefore, even at the $K\to 0$ limit (or $m_e\to \infty$ limit), we do not need to impose the conservation of ${\tl Q}$.

\paragraph{Usual axion-photon conversion}
We also comment that the symmetry argument can apply to conventional axion-photon conversion with helical undulator-like magnetic field configurations. In the `vacuum' with only this stationary magnetic field, we have symmetry corresponding to $\hat{H}$ and 
$\hat Q'=\hat{P}^z+k \hat{L}^z$. Assuming that we have the initial photon propagating in the $z$-direction with the spin $s^z$, we have 
$w_\g=w_\f$, and $k_{\f}^z=k_{\g}^z+k s^z$ from the $H$,  and $ Q'$ conservation, respectively. Then we get $\sqrt{w^2_{\f}-k_{\g,x}^2-k_{\g,y}^2}-\sqrt{w_\f^2-k_{\g,x}^2-k_{\g,y}^2-m_\f^2}=-k s^z$, where we also used momentum conservations in the $x,y$ directions. Thus this cannot be satisfied with $s^z=-1$ and $m_\f^2\simeq 2 k w_\g$, which is nothing but the resonance condition~\cite{Abel:2006qt,Arias:2010bh,Bahre:2013ywa,Ortiz:2020tgs}. We note that this can happen with $s^z=1$.\footnote{Thus, the photon produced by a helical undulator cannot be efficiently converted to the ALP due to another aligned helical undulator in the front if they have the same helicity.} This is relevant to why the on-shell photon production followed by the axion-photon conversion in the same undulator is suppressed in the $z$-direction for any $m_\f$ in our case. The conversion occurs via the boundary effect, which gives the amplitude of axion-photon conversion $\propto g_{\f\g\g}B_0L\times (kL)^{-1}.$ Since the photon number in the $z$-direction is produced with $\propto K^2 e^2 k L $, we get the number of produced ALP, $\propto g_{\f\g\g}^2B_0^2 K^2 e^2 L/k,$ which is smaller than the contribution discussed in the main part.
\\
\\
Although we mostly focused on the production of the ALP, the symmetry argument can easily be applied to a generic final state particle. 
Due to the symmetry, the spin of the final state particle and the orbital angular momentum, and thus the angular dependence of the particle’s motion, are related. This means that we may measure the spin of the final state particle by precisely measuring the orbital angular distribution of the particle.

\section{Alternative Estimation for ALP Production from Background Fields}
\lac{nume}
To show the validity of the previous analytic results, let us perform another estimation from a different strategy. 
We treat not only $\vec{B}_{\rm ext}$ but also $\vec{E}$ from the electron radiation from \Eq{ele} background fields and estimate the ALP production numerically.\footnote{We can also perform a semi-analytic estimation as before by moving to the rest frame of the electron by again using $\d_{L/(\b^z\g)}$ etc in the relevant Fourier transformation. Interestingly, with $K\ll1$, in this frame, the magnetic field behaves like an electromagnetic wave, and we get a similar picture to the usual axion-photon conversion, but the stationary external field is the electron's Coulomb potential. We have checked that the agreement of the analytic result with \Eq{ana}.} Namely, the electron is integrated out. 
With the background $\vec E$ and $\vec B$, the amplitude of the axion-photon conversion can be estimated as
\begin{align}
\laq{ampnum}
\langle\f,k_\f| 0_{E,B} \rangle= 
-i \int{d^4x \frac{g_{\f\g\g}}{\sqrt{2w_\f}}e^{i\(-\vec{k}_\f \cdot \vec{x} + w_\f t\)} \vec{E} \cdot \vec{B}_{\rm ext}}, 
\end{align}
where we omit the magnetic field radiated from the electron which is perpendicular to the electric field. 
 Then, we can estimate the number of the ALPs by
\beq
\laq{Nfull}
n_{\f}=\int \frac{d^3 \vec k_{\f}}{(2\pi)^3}\ab{\langle\f,k_\f| 0_{E,B} \rangle}^2
\eeq
Since we know both $\vec{B}_{\rm ext}$ and $\vec E$ from \Eq{ele}, we can estimate the amplitude by performing the integration numerically.

The advantage of this analysis is that we do not need to rely on the $K$ expansion, and we can also check the case that $L$ is not very large as well as more generic setups. The disadvantage is the high numerical costs with the integrals with multi-variables, treating which with an economical method is beyond our scope, but the strategy of the calculation is straightforward.\footnote{We consider that the estimation should be possible to be incorporated in the software {such as SPECTRA}~\cite{Tanaka:il5065} for calculating the synchrotron radiation emitted from an undulator. }

In the integration of \Eq{ampnum}, we change the time variable with $d t =d t_e\times \frac{dt}{dt_e}$ where $t_e$ given in \Eq{terela}. For the 3D space, we use cylinder coordinates along the $z$-direction: $(x,y,z)=(r\cos\Phi,r\sin\Phi,z)$. For numerical convenience and also to match the realistic case, we set a cutoff for the integration of $r=[0-r_{\rm cutoff}]$. 
The integration range of \Eq{ampnum} is set as $t_e=[0-L/\b^z].$\footnote{Strictly speaking, for the Coulomb potential contribution, $t_e<0$ is also relevant. We checked by including the contribution, the result does not change much.}  

\begin{figure}[t!]
    \begin{center}  
        \includegraphics[width=145mm]{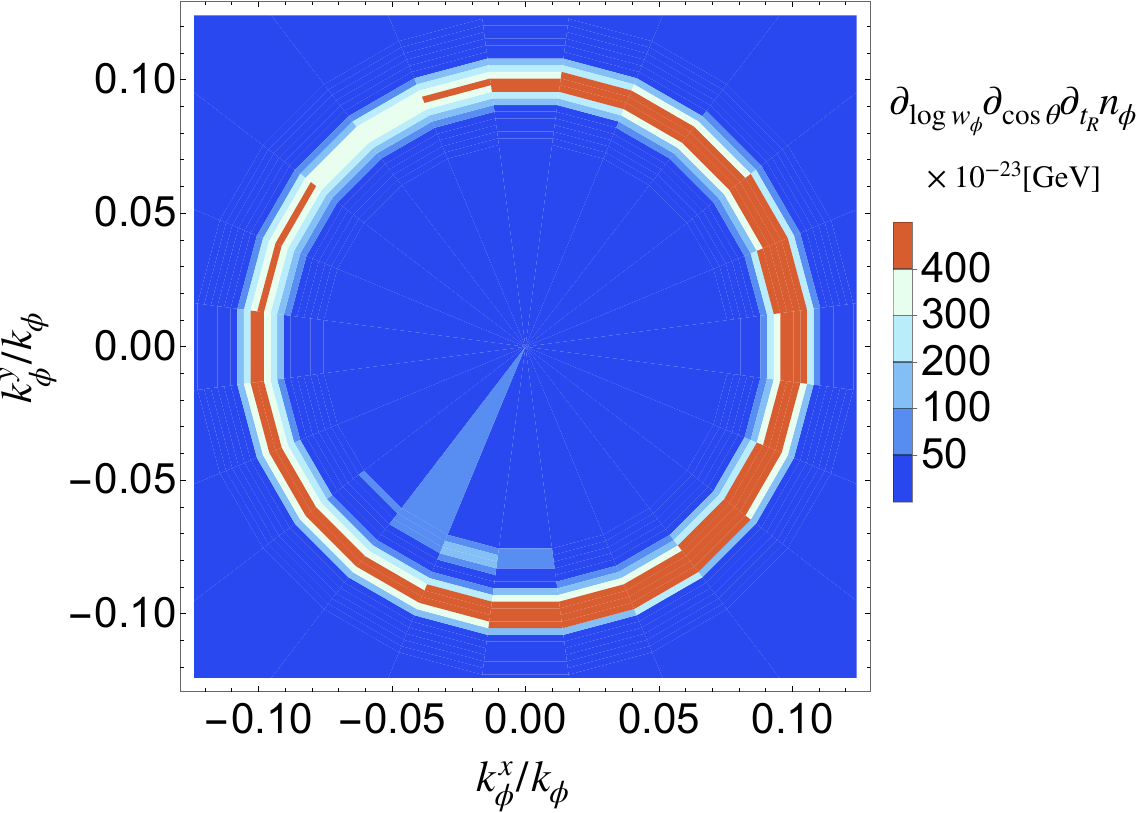}
    \end{center}
    \caption{The contour plot of $\partial_{w_\f}\partial_{\cos\theta}\partial_{t_R}n_\f$ in the $k_{\f}^x/k_\f, k_{\f}^y/k_\f$ plane.  We fix $k_\f=k \g^2$, and set $\gamma=10$, $2\pi/k= 0.5$mm, $B_0=1$Tesla, $L=8\times 2\pi/k$ and $m_\f=0$.  $m_e$ and $e$ are taken to be the realistic values. 
 This corresponds to $K\approx 0.05.$ }
    \label{fig:contour2}
    \end{figure}

\paragraph{Simulations and consistencies with light ALP}
We set $m_\f=0, \k=1,\phi=0, r_{\rm cutoff}\approx \pi/k, L=5  \times 2\pi/k$ and $\g=10$. The numerical result of the ALP differential flux (defined by $t_R^{-1}\partial_{\cos\theta}\partial_{\phi}n_\f$) in the $k_{\f}^x,k_{\f}^y$ plane is shown in Fig. \ref{fig:contour2}. Here, $K\approx 0.05$ which is much smaller than unity. 
With $k_{\f}^x=k_{\f}^y=0$, the production is suppressed, while it is produced dominantly in the direction of $\theta \sim 1/\g$.
We see it is mostly independent of $\tl \f$.\footnote{The slight dependence on $\tl \f$ gets more significant when we use smaller $L$, implying that this is a boundary effect.}

\begin{figure}[t!]
    \begin{center}  
    \includegraphics[width=145mm]{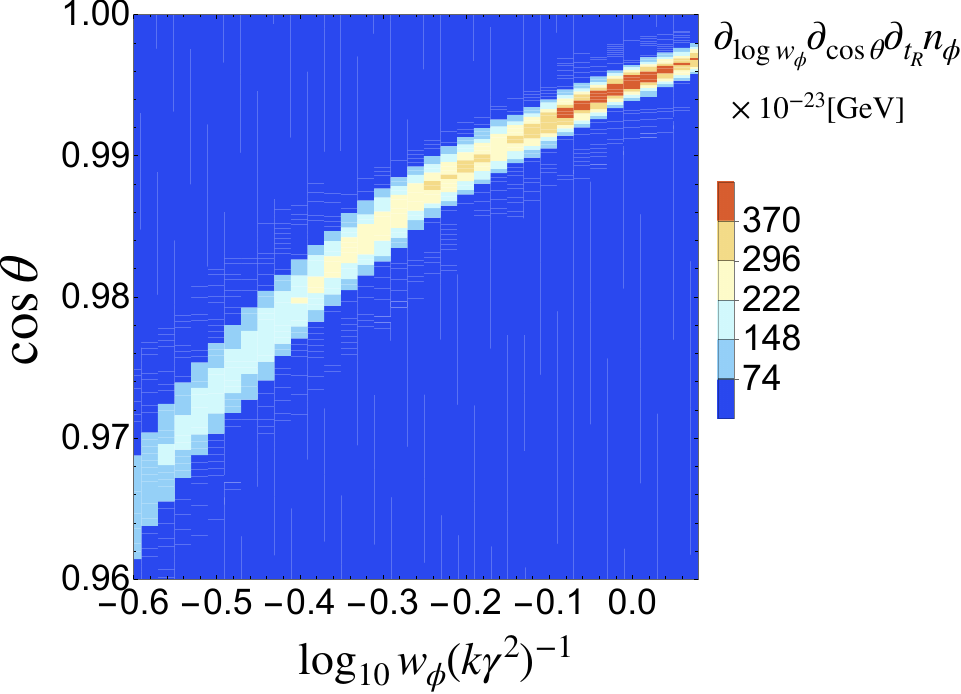}
    \end{center}
    \caption{The contour plot of $\partial_{w_\f}\partial_{\cos\theta}\partial_{t_R}n_\f$ in the $\log_{10}{w_\f (k\g^2)^{-1}}, \cos\theta$ plane.    We set $\gamma=10$, $2\pi/k= 0.5$mm, $B_0=1$Tesla, $L=5\times 2\pi/k$ and 
 $m_\f=0$. 
 $m_e$ and $e$ are taken to be the realistic values. 
 This corresponds to $K\approx 0.05.$ 
 }
    \label{fig:contour1}
    \end{figure}

In Fig.~\ref{fig:contour1}, we show the differential flux in $w_\f.$ 
Instead of performing $\tl{\f}$ integral, we set $\tl \f=0$ and multiply $2\pi$ in the final result to reduce the calculation cost. 
The relationship is consistent with \Eq{ana}.
The finite width is found to be consistent with the violation of the energy-momentum conservation due to the `delta-function', $\frac{\D k_{\f}}{k_\f}\sim  \frac{\pi }{L k}$. We have confirmed that the dominant contribution indeed comes from the first term of \Eq{ele} with the highly suppressed second term.

\begin{figure}[t!]
    \begin{center}  
        \includegraphics[width=145mm]{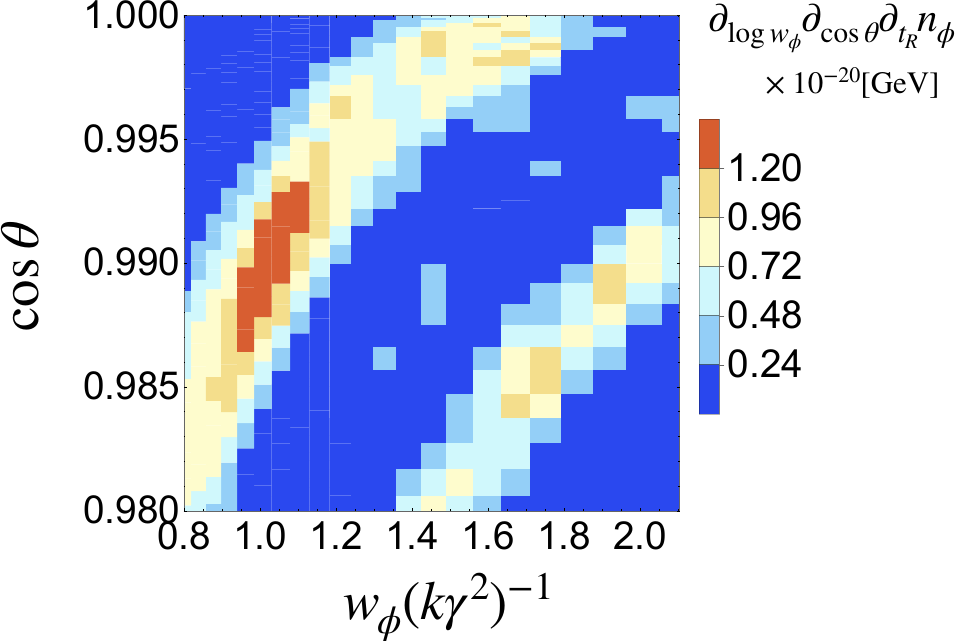}
    \end{center}
    \caption{The contour plot of $\partial_{w_\f}\partial_{\cos\theta}\partial_{t_R}n_\f$ in the ${w_\f (k\g^2)^{-1}}, \cos\theta$ plane.  We set $\gamma=6$, $ 2\pi/k= 5$mm, $B_0=1$Tesla, $L=3\times 2\pi/k$ and $m_\f=0 $. 
 $m_e$ and $e$ are taken to be the realistic values. 
 This corresponds to $K\approx 0.5$.
    }
    \label{fig:contour3}
    \end{figure}

In Fig.~\ref{fig:contour3}, we show the result for a larger $K$ with $K=0.5$. In this case, $\O(K^2)$ terms also contribute. 
We set $m_\f=0, \k=1,\phi=0, r_{\rm cutoff}\approx\pi/k, L=3 \l.$ The result is consistent with our analytic estimate, including the presence of the second peak at a larger $w_\f.$

\begin{figure}[t!]
    \begin{center}  
        \includegraphics[width=145mm]{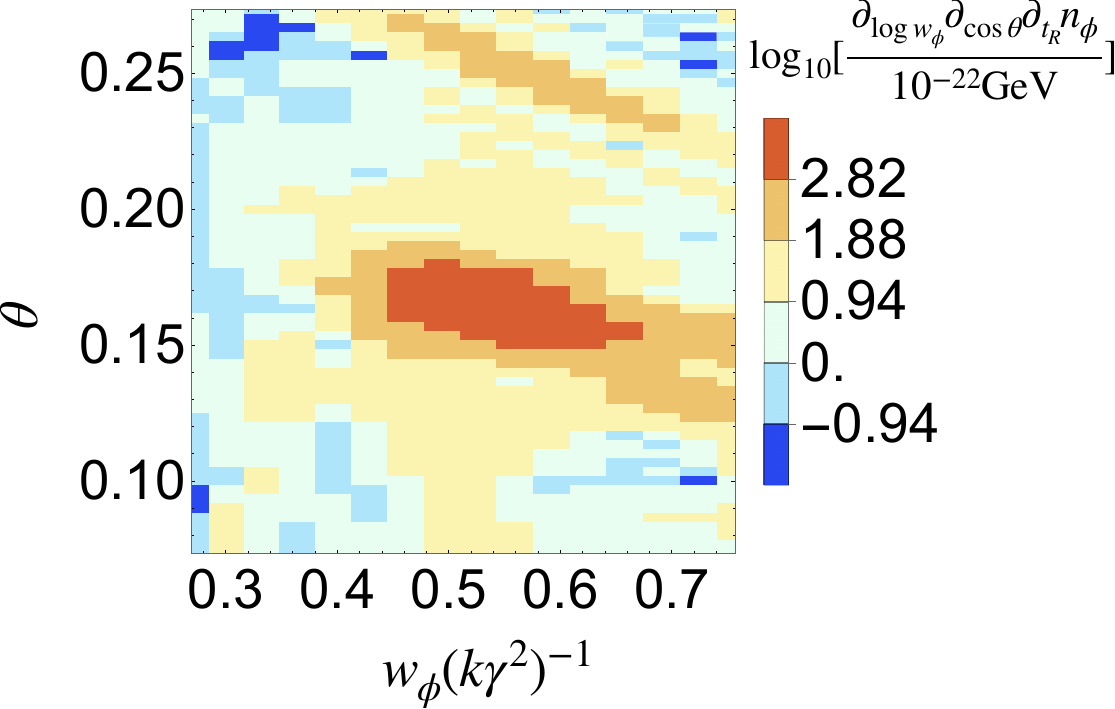}
    \end{center}
    \caption{The contour plot of $\log_{10}[\partial_{w_\f}\partial_{\cos\theta}\partial_{t_R}n_\f]$  in ${w_\f (k\g^2)^{-1}}, \theta$ plane, with $m_\phi=k \g.$ 
We set $\gamma=10$, $2\pi/k= 5$mm, $B_0=1$Tesla, $L=6\times 2\pi/k$, $m_\f=\g k$. 
 $m_e,e$ are taken to be the realistic values. 
 This corresponds to $K\approx 0.5$.
    }
    \label{fig:reso}
    \end{figure}

\paragraph{Simulations and consistencies in the  resonance regime}
In Fig.\ref{fig:reso}, we show the case with $K=0.5$ with massive ALP to check the resonance behavior. We take $m_\f=k \g$, $\k=1,\phi=0, r_{\rm cutoff}= \pi/k, L=6 \times 2\pi/k.$ 
The result is consistent with our analytical estimate for the resonance condition~\eq{resopeak}, which corresponds to the center of the figure. 
The center of the figure has a number flux of 2-3 orders of magnitudes larger than the other places, agreeing with the analytical estimation: $\sim \frac{\pi (\d_L(0))^4}{(\d_L(k))^4} \sim \(\frac{L k}{2\pi}\)^4 \pi =\O(10^{2-3})$. The scaling with $L^4$ is also checked. 
\\

In all previous figures, the flux values are consistent with the analytical estimates. Based on these results, we conclude that our analytical estimation in \Sec{ALP} has indeed captured the physics of ALP production in undulators.

\bibliography{undulator3.bib}

\end{document}